# Source localization of the EEG human brainwaves activities via all the different mother wavelets families for stationary wavelet transform decomposition.

Tarek Frikha[2]
[2]CES-Laboratory, Sfax University, National Engineering School of Sfax, Tunisia
tarek.frikha@enis.tn

*Abstract*— The source localization of the human brain activities is an important resource for the recognition of cognitive state, medical disorders and a better understanding of the brain in general. In this study, we have compared 51 mother wavelets from 7 different wavelet families in a Stationary Wavelet transform (SWT) decomposition of an EEG signal. This process includes Haar, Symlets, Daubechies, Coiflets, Discrete Meyer, Biorthogonal and reverse Biorthogonal wavelet families in extracting five different brainwave sub-bands for a source localization. For this process, we used the Independent Component Analysis (ICA) for feature extraction followed by the Boundary Element Model (BEM) and the Equivalent Current Dipole (ECD) for the forward and inverse problem solutions. The evaluation results in investigating the optimal mother wavelet for source localization eventually identified the sym 20 mother wavelet as the best choice followed by bior 6.8 and coif 5.

## I. Introduction

To accurately study and analyze the human brain, the Electroencephalography (EEG) [1] is thought to be the optimal method that helps us advance in our quest due to the non-invasiveness and the low-cost factors. In fact, the EEG provides access to human brain activities in 5 main frequency band packages presented in table I [2]. The scientific trend shifted towards exploiting these frequency sub-bands and seeking the extraction of pure and non-contaminated signals instead of developing recording methods and other ways to express the signal.

The research community took omnidirectional approaches throughout the recent years to try to extract the human brain activities and access these five different frequency sub-bands. In this context, Murali, L., et al [3] used the recurrence quantification analysis (RQA) algorithm and an adaptive FIR filter for the EEG signal extraction. As for Singh, Vivek, et al [4], they compared using Finite Impulse Response (FIR) and Infinite Impulse Response (IIR) filters and confirmed the FIR success over RII regarding EEG signals. On the other hand, Nallamothu, S., et al [5] used a Nonlinear Least Mean square (LMS) adaptive filtering to remove artifacts from EEG signal.

For their part, Tzimourta, K. D., et al [6] used a Discrete Wavelet Transform (DWT) for feature extraction and Support Vector Machine (SVM) classification of Epileptic Seizures. Actually, our previous Work [7] has proved the effectiveness of the Stationary Wavelet Transform (SWT) using Symlet 4 mother wavelet compared to FIR filters in feature extraction of the alpha and gamma band waves. Furthermore, Akkar, H. A., et al [8] proved that the Symlet 9 mother wavelet is the best wavelet from a set of 25 mother wavelet functions using the Packet Wavelet Transform (PWT). Lema-Condo, Efrén L., et al [9] compared 18 different mother wavelets for EEG signal analysis and affirmed Symlet 6 and Daubechie 5 are the most adequate for EEG signals. Al-Qazzaz, Noor, et al [10] compared 45 mother wavelets to conclude that Symlet 9 followed by Coiflet 3 and Daubechie 7 exhibits the highest similarities and compatibilities with the EEG signal after applying an FIR notch filter.

TABLE I. THE EEG SIGNAL MAIN HUMAN BRAIN FREQUENCIES.

| EEG bands | | Frequency bands (Hz) | Main description |
|---|---|---|---|
| Delta | | 0-4 | Deep state of sleep |
| Theta | | 4-8 | Deep meditation & lucid dreaming |
| Alpha | | 8-12 | Relaxation/Creativity |
| Beta | | 12-32 | Analytical thinking or Stress/Anxiety |
| Gamma | lower | 32-64 | Wide brain activities or brain disorder |
| | Higher | 64-128 | |

The EEG signal is a non-stationary signal, the advantage of using the Wavelet transform over the usual Fourier transform in EEG signals is their capability to analyze non-stationary signals [11] due to their improved presentation in both the time and frequency domain as shown by figure 1.

In this context, this study aims at comparing 51 different mother wavelets using SWT to extract human brainwaves and localize their sources. In section II, we will address the methodology first, by describing the manipulated dataset and then, proceed by presenting the SWT and the processing steps of our study, and finally by introducing the evaluation methods. Section III will feature the conceived results and section IV will highlight the discussion.

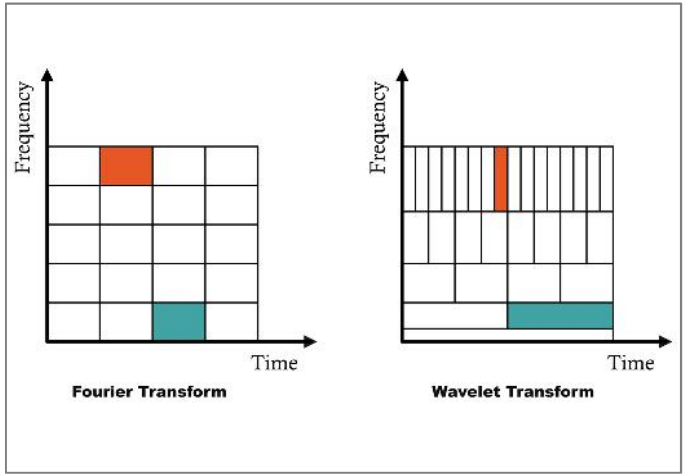

Figure 1.Comparision between the partition of Fourier Transform and Wavelet Transform in time-frequency domain.

II. METHEDOLOGY

*A. Dataset*

*1) Simulated signal*

Influenced by the morphology and the structure of actual EEG signals, we created a sinusoidal signal with oscillations of 400ms on 800ms time windows used in the evaluation process. A sampling rate of 1000 Hz and an oscillation frequency of 3, 6, 10, 20, 45 Hz were recorded for the extraction of Delta, Theta, Alpha, Beta, and Gamma waves, respectively. The signal to the noise ratio (SNR) was also altered from -5 to 15 dB with -5 dB for noisy signal simulation, 10 dB for balanced signals, and 15 dB for acceptable quality signals. On the other hand, the amplitude of the signal depends on both the SNR value and the noise contaminating the signal, which is what is known as a pink noise, besides, it is very common noise for biological systems.

*2) The EEG dataset*

The EEG signal dataset used in this study is a-one subject recording of a pre-surgical EEG signal from a pharmaco-resistant subject with a symptomatic focal cortical dysplasia in the right occipital-temporal junction. The acquisition and preprocessing phases were applied as in our previous work [7] and validated by an expert neurologist. This particular EEG recording was chosen because it presented clear alpha and gamma patterns with regular spiking and visible epileptic oscillations as validated by the expert. The EEG data was recorded on a Deltamed System, with a 2500 Hz sampling rate and anti-aliasing low-pass analog filter set to 100 Hz. The dataset contained 74 epochs with a 6-second duration each, 62 channels and 148 events.

*B. The Wavelet Transform*

Similar to the Fourier Transform (FT), the Wavelet Transform (WT) is a function that grants the passage from the time to the frequency domain. However, the FT decomposes the signal into a series of sinus and cosines components as in the Equation (1):

$$s(t) = \int_{\omega=-\infty}^{+\infty} S(\omega)e^{j\omega t}\,d\omega \qquad (1)$$

with S (ω) the Short-Time Fourier coefficient controlled using the frequency parameter ω.

The Wavelet Transform also decomposes the signal into a series of wavelet components as in equation (2):

$$s(t) = \int_{a=0}^{+\infty}\int_{b=0}^{+\infty} C(a,b)\varphi_{a,b}(t)\,da\,db \qquad (2)$$

where C (a, b) is the wavelet coefficient and φa,b(t), the mother wavelet with "a" as the scaling parameter and "b" the wavelet shifting parameter that determines the shape of the wavelet. In fact, figure 2 highlights the difference between FT and WT decomposition components. Moreover, the wavelets are characterized by a limited duration, irregularity and asymmetricity compared to the predictable, fluid, infinitely propagated sinus waveform.

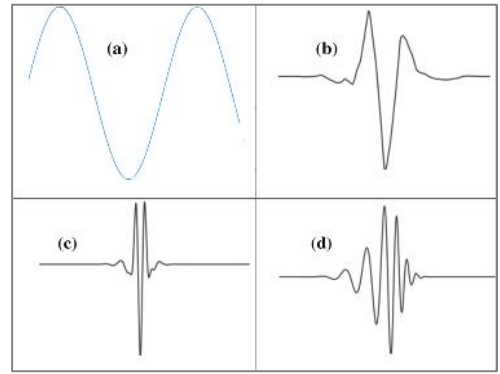

Figure 2.Comparision between (a) FT decomposition component and different mother wavelets families decomposition components (b) symlets 4 (c) coiflets 5 (d) daubechies 11.

On the other hand, the Wavelet Transform used in this study is the Stationary one (SWT) instead of the Continuous Wavelet Transform (CWT) or the Discrete Wavelet Transform (DWT). In fact, the SWT is more suitable for our case by avoiding the frequency band overlapping of CWT [12] and preserving the properties of the signal by averting the binary decimation process (down sampling) of DWT [13].

*C. Levels of Decompositions and processing steps*

In order to decompose the EEG signal of our dataset that has 2500 Hz sampling rate to extract the five EEG frequency sub-bands, we had to reduce the signal to exactly 2048 Hz sampling rate otherwise, these sub-bands would be extremely overlapping. In figure 3, we display the decomposition of the resampled EEG signal. We notice here that in our previous study [7], we have not resampled the signal as we extracted only the alpha and gamma waves that were far separated and did not cause band overlapping issues. Our decomposition Level was 9 to acquire the access to the delta wave frequencies while our previous work needed only 7 levels of decomposition to reach the alpha wave. We can also notice that in our previous work, the Approximated coefficients cAi included upper and lower levels (for alpha wave extraction the cAi were 6,7,8 and cDi was 7), while for this study, we have included only the above upper levels for the cAi (for alpha wave extraction the cAi were 6,7 and cDi was 7).

As the wavelet decomposition phase is completed, we evaluate the mother wavelets used in this process and move on to the source localization. Figure 4 shows the processing steps of this study.

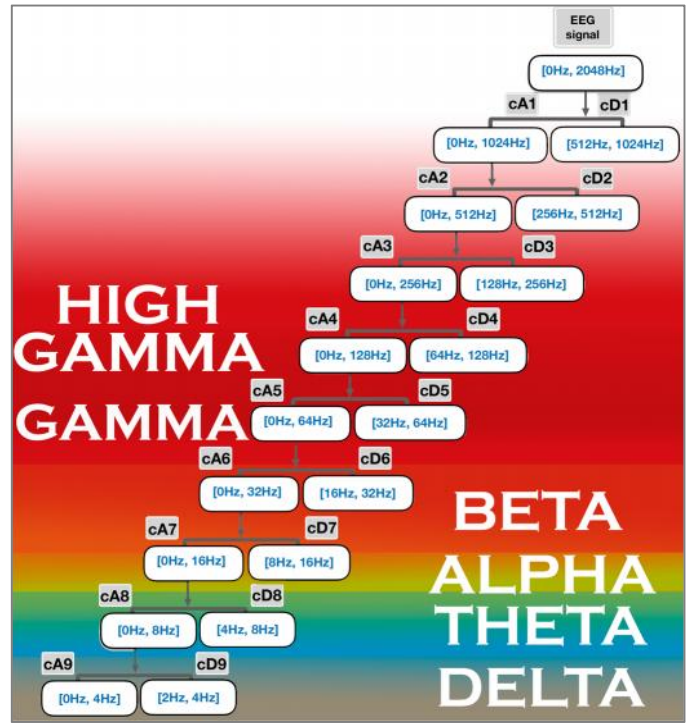

Figure 3. EEG signal SWT decomposition levels with cAi as the Approximated coefficients and cDi the detailed coefficients.

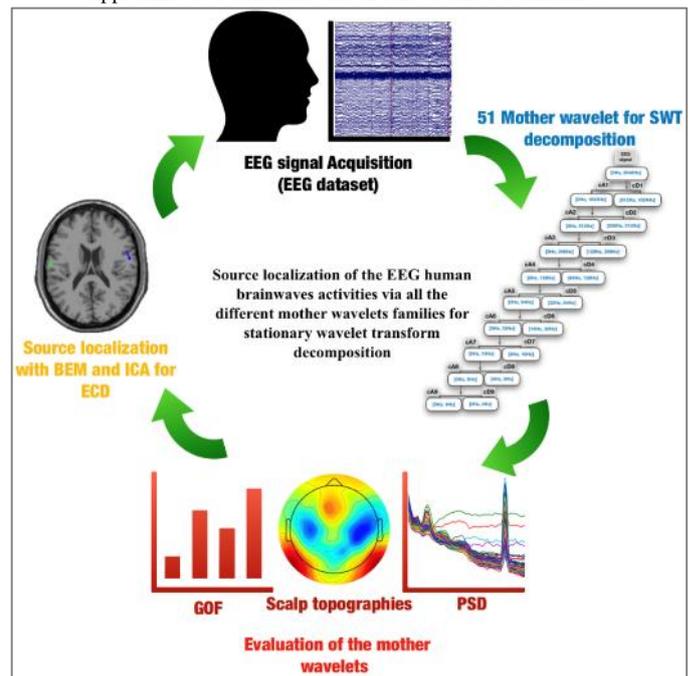

Figure 4. The cycle of processing steps during this study.

*D. The Evaluation methods*

  *1) The Goodness Of Fit (GOF)*

The Goodness Of Fit (GOF) is an evaluation method commonly used for physiological signals that adopt Pearson's chi-squared statistical test [14], which is the normalized sum of squared deviations that investigate the likelihood of an observed

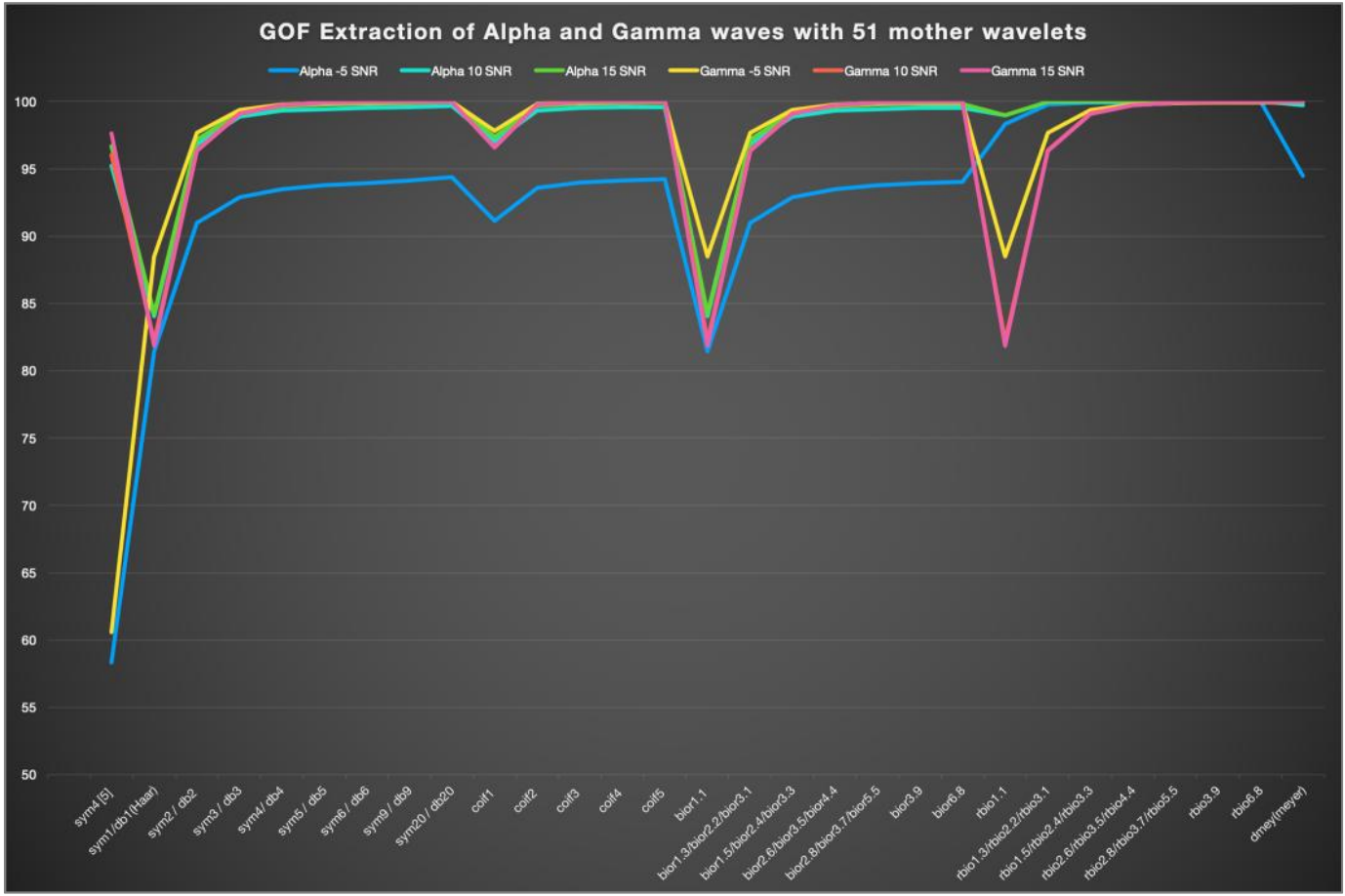

Figure 5. The GOF results in alpha and gamma waves with 51 mother wavelets using SNR values of -5,10 and 15dB.

difference in the frequency distribution compared to the theoretical distribution as in equation (3):

$$G = 1 - \left(\frac{\sum_{t=1}^{r}\bigl(s(t)-s_f(t)\bigr)^2}{\sum_{t=1}^{r}s(t)^2}\right) \quad (3)$$

where s(t) is the theoretical power and sf(t) the power of the extracted signal that depends on the adopted mother wavelet.

*2) The Power Spectral Density (PSD) and scalp topographies*

The Power Spectral Density (PSD) is a display of the data energy distribution throughout the frequency spectrum. It is used as a visual evaluation process for its efficiency in presenting the data in the frequency domain rather than the time domain, which allows the identification of the extracted EEG frequency bands [15]. The energy frequency distribution of the EEG signal channels compares the mother wavelets effectiveness in isolating the extracted frequency band from the other sub-bands or artifacts and differentiates its capabilities to amplify the extracted signal power.

On the other hand, the scalp topographies are another visual evaluation process since it represents a mapping of the brain activities distributed on the surface of the scalp. An increasingly dipolar topography suggests that a cerebral measurement is an observation of a discharge operation involving a big number of neurons. Even in non-epileptic observations of brain activities, the dipolar scalp topographies are a great indicator of a valuable recording session since they reflect the domination of certain areas over others in the energetic exertion, which is the typical and more natural habit of cerebral behavior [16].

*3) The source Localization*

The source localization is an estimation of the brain activity generator locations [17]. To reach this estimation, first, we solve the forward problem, which is a calculation of the field generated by a given source for an estimated brain shape and conductivity, with a consideration of numerous properties, such as the shape of the brain that changes from a subject to another or the anisotropy conductivity of the skull and the brain conductivity [18]. For the forward problem, we used the Boundary Element Model (BEM), which is a surface mesh calculation of interfaces between the tissues using the MRI of the patient (which makes it a realistic model) [19]. For the inverse problem, which is an estimation of the current generator distribution responsible for the electric EEG signal, we used the equivalent current dipole (ECD), which is the most used method to simplify the brain activities in a few sources [20]. The signals are assumed to be generated by a small number of focal sources modeled by current dipoles (an unknown position, amplitude and orientation). Moreover, the extracted signal has to undergo an independent component analysis (ICA) dipole fitting operation as a preprocessing phase before the ECD inverse problem solution, in order to separate different components and make the components in a dipolar state useful in the localization of the source generators. The ICA is the feature extraction phase compatible with the statistically independent and non-Gaussian signals, which are the traits of the EEG signal [21] while the ECD and the BEM are our classification algorithm.

In fact, the source localization process is sensitive to the quality of the extracted EEG frequency band and can also serve as an evaluation process that depends on the number of the located sources and the accuracy of their localization.

## III. RESULTS

### A. The GOF evaluation results

The goodness of fit (GOF) is the evaluation process that enabled us to minimize both our wavelet selection and processing criteria. Considering that the other evaluation methods and the source localization are a computationally heavy and costly process, the GOF is an excellent fast evaluation that relieved us from repeating the hull processing steps and source localization for the vast number of 51 mother wavelets. Figure 5 presents the GOF results for the 51 mother wavelets with different SNR values of -5, 10 and 15 dB as we have mentioned in the dataset descriptions in section II, A, 1).

On the other hand, the use of alpha and gamma wave extraction in GOF evaluation is justified by our earlier knowledge during our previous study [7] of the excellent capability of SWT in extracting these specific frequency sub-bands.

The GOF results showed a similar pattern across the different frequency sub-bands and different SNR values with a distinct superiority to sym20, coif5, bior6.8, rbio6.8 and dmey wavelets.

In order to explore and investigate this superiority, we have extracted the best mother wavelets of every wavelet family and the wavelets that already showed some noteworthy results in other studies, such as sym4 in [7], db5 and sym6 in [9] and sym9 in [8] [10], in every EEG frequency sub-band, as shown in figure 6.

Besides, after isolating the GOF results about the limited number of noteworthy wavelets, we notice that the performance of the wavelet extraction changes from frequency sub-band to another with an obvious preeminence in gamma and alpha waves. We also observe that sym4 in [7] is the lowest in the GOF performance due to the Approximated coefficient choice in the decomposition phase compared to our choice of Approximated coefficient in this study for all the wavelets.

Finally, to lock the GOF evaluation results, we calculated the noteworthy wavelet average across the five frequency sub-bands and ordered them from the lowest performance, on the left, to the best performance, on the right by their GOF score in figure 7. In fact, the best results were achieved using demy and sym20 wavelets, while the worst results used sym4 [7] and Haar also.

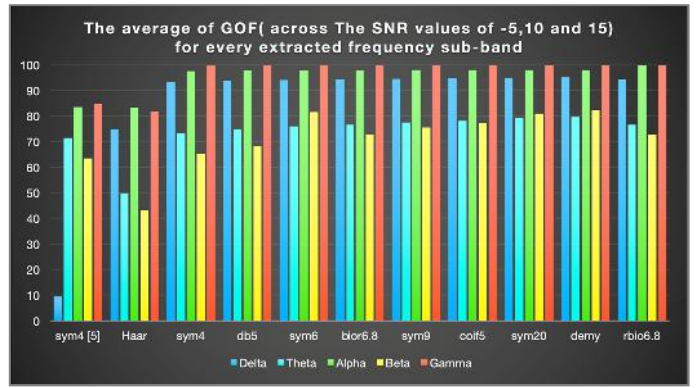

Figure 6. The results of the average GOF for every EEG frequency sub-band with the selected mother wavelets across the SNR values of -5,10 and 15dB.

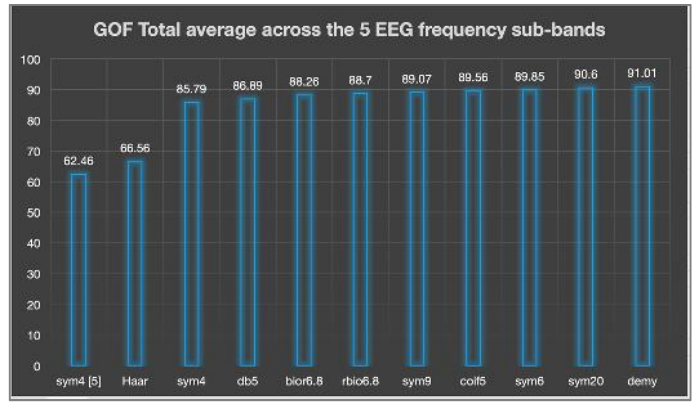

Figure 7. The GOF average of the noteworthy wavelets across the EEG frequency sub-bands and SNR values ranked from left to right by order of best performance.

### B. The PSD and topographies evaluation results

The Power Spectral Density (PSD) is also an important evaluation method that grants us a visual representation of the EEG signal extraction. The choice of frequency sub-band extraction visualization for this evaluation was limited to the alpha and gamma waves for the confirmed potential of SWT in their extraction. Moreover, due to the weak energy of the gamma wave and its proximity to the 50 Hz noise artifact of the original EEG signal dataset, we relied only on the alpha wave in the PSD visualization as it provides a clear display of the extraction effectiveness difference between the selected mother wavelets.

In Figure 8, we compare the EEG signal extraction of the alpha frequency sub-band using the different noteworthy wavelets chosen by the GOF evaluation ordered from the worst to the best. As we can deduce, the haar and sym4 wavelets respectively had the worst results with a signal spectrum contaminated by different artifacts and other frequency sub-bands while sym20 and dmey had the best results in isolating the extracted signals from other infiltrating ones. We can also recognize the abilities of the new SWT decomposition in eliminating high frequency, while witnessing some difficulties in low frequency elimination, such as delta and theta, as demonstrated in the PSD visualization.

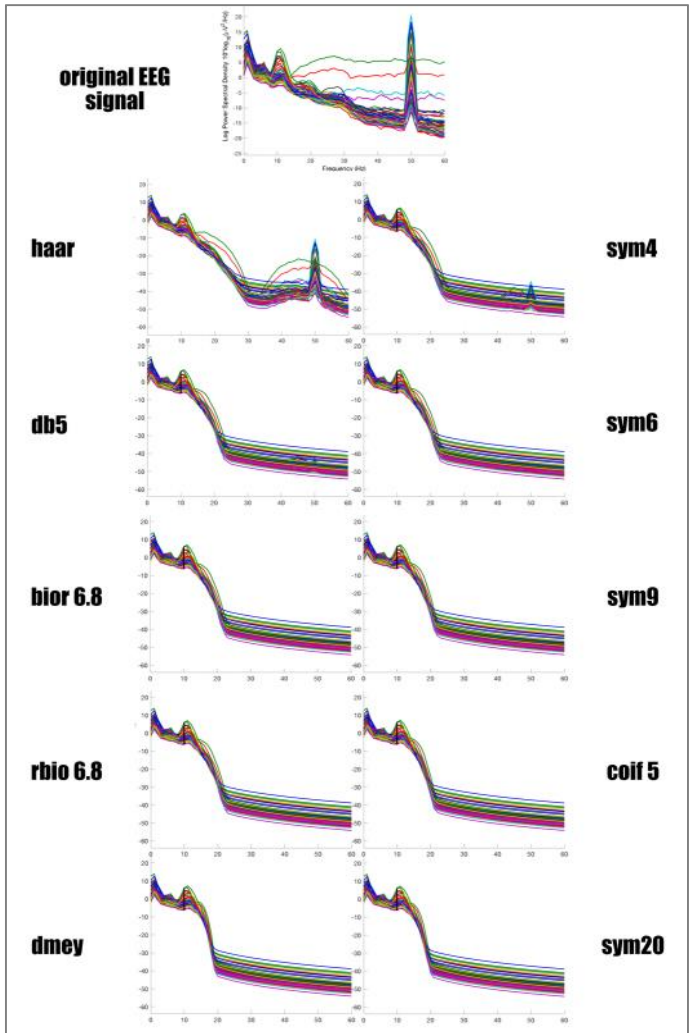

Figure 8. PSD Visualization of the different noteworthy mother wavelets in alpha wave extraction.

For the scalp topography visualization, almost all the noteworthy mother wavelets selected by the GOF had similar good results by producing depolarized scalp topographies isolated from the other frequencies, except for the Haar and sym4 wavelet extractions, which produced some interfering artifacts that could compromise the ability of reviewing the scalp topographies by the medical experts and mislead them in diagnosing the cause of these parasites. Figure 9 displays the scalp topographies of the original signal compared to both the mother wavelet extraction and the contaminated scalp topographies of haar and sym4. As an assessment of the PSD and scalp topography evaluation, the sym20 and demy mother wavelets demonstrated the best results while the haar and sym 4 produced the worst ones.

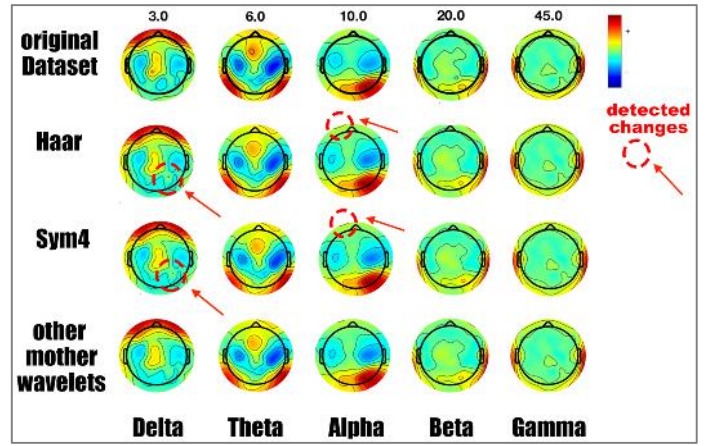

Figure 9. A scalp topographies comparison between the original dataset, the noteworthy mother wavelets extractions and the contaminated scalp topographies of haar and sym4 wavelets for the five EEG frequencies sub-bands.

## C. The source localization

For the source localization, we preformed the Independent Component Analysis (ICA) on the extracted signals by the noteworthy mother wavelets, then we used the BEM for the forward problem and ECD for the inverse problem. As we have already mentioned, the ICA is a computationally costly process for feature extraction, especially with 62 EEG channels for the extraction of the same number of components before the source localization, so we reduced the process to include only the alpha and gamma frequency sub-bands. The alpha wave is the most important brainwave activity in the human brain and the gamma wave is perceived as an indicator of high active cognitive state and constantly used in brain malfunction and disease confirmation [22].

The ICA was performed using the runica algorithm from the EEGLAB toolbox [23]. Then, the BEM and ECD were executed using the fieldtrip toolbox [24]. We set a rejection threshold for the components based on the Residue Variance equivalent to RV=15% as it is the optimum value in component rejection, as confirmed by Artoni et al [25]. In figure 10, we present the source localization of the alpha and gamma extracted waves using the different noteworthy mother wavelets. As we can see, every mother wavelet extraction has a different number of sources localized under the Residue Variance RV error threshold and different source locations compared to each other. In order to evaluate the source localization of our different mother wavelets, we focus on the number of localized components by every mother wavelet and the number of times every mother wavelet has the best accuracy (lower RV value) in localizing the source of a component and the average of accuracy in the five first components. The reason for which we have included the accuracy of the five first components in our evaluation is that the ICA using the runica algorithm for the output components in a decreasing order of the EEG variance accounted for by each component, i.e. the lower the order of a component, the more data (neural and/or artifactual) it accounts for [26].

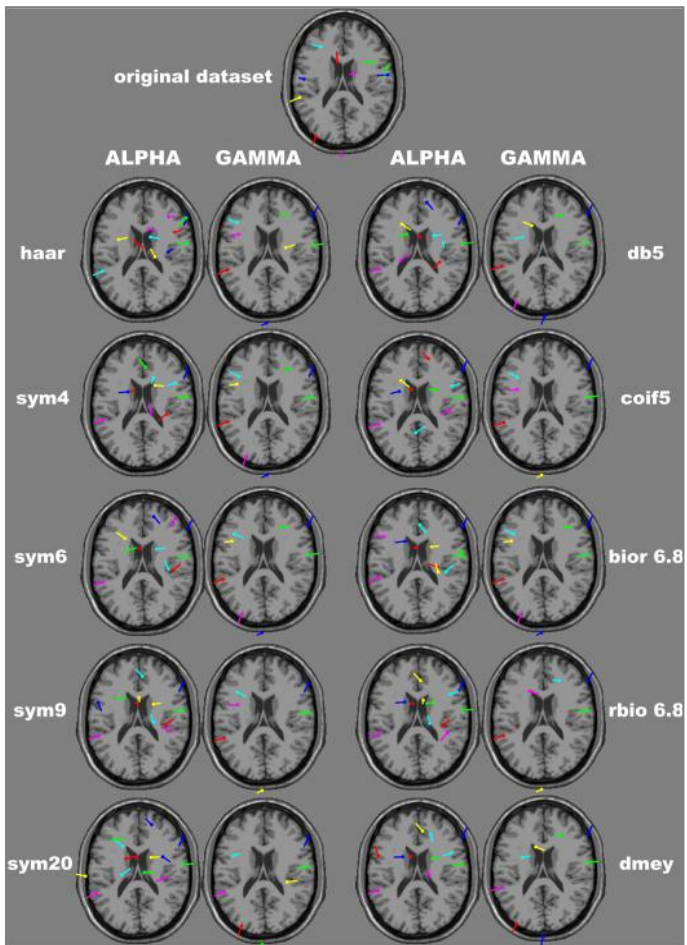

Figure 10. Visualization of the alpha and gamma waves source localization using the noteworthy mother wavelets extractions.

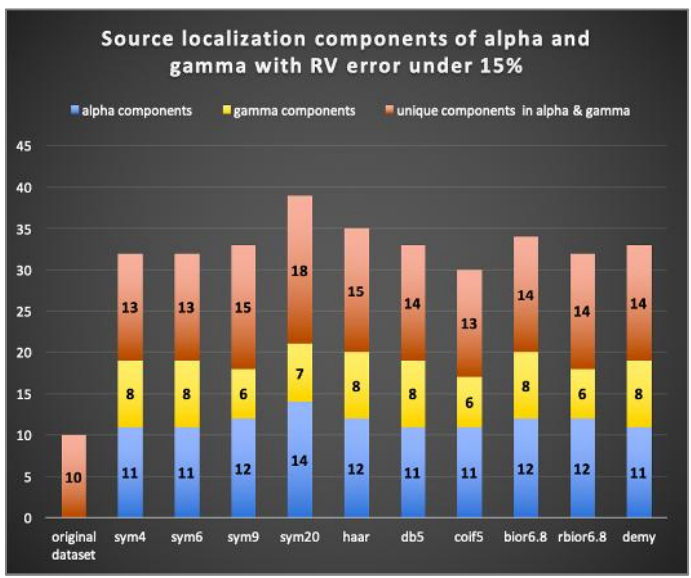

Figure 11. The number of component localized by the noteworthy mother wavelets in alpha and gamma waves with RV under 15%.

Figure 11, shows the number of components localized by each mother wavelet in the alpha and gamma frequency sub-bands and only the number of components that were not localized in the other frequency sub-bands.

An interpretation of the number of localized component results showed that the sym20 mother wavelets produced the best results followed by Haar and bior6.8, while coif5 had the lowest number of localized components.

In Figure 12, we explore the accuracy of the noteworthy mother wavelets in source localization by comparing the number of times each mother wavelet managed to record the lowest RV score. This chart also considers the localization in the alpha wave, gamma wave and in the combined best localized components of both frequency sub-bands.

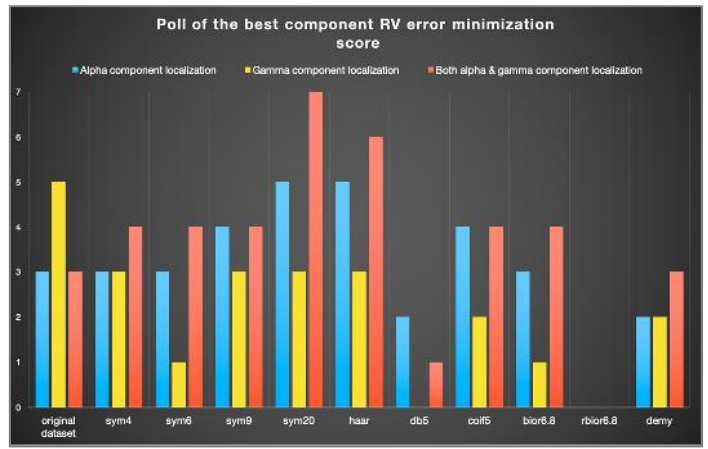

Figure 12. The number of times each mother wavelet scored the best accuracy for source localization in alpha and gamma waves.

The sym 20 mother wavelet scored the best accuracy results followed by haar and sym9, while rbio 6.8 did not have even once the best accuracy compared to the other wavelets for both frequency sub-bands. We also spot that the original EEG signal had an impressive accuracy in gamma wave, which indicates the interference of the other frequency sub-bands or the 50Hz noise artifact and compromised the integrity of the located sources considering that the gamma wave had poor frequency energy that could not produce such result.

Table II presents the final criteria for source localization evaluation that focuses on the accuracy of the five first components. The accuracy is expressed with the RV values, which means that the lower the RV value is, the better accuracy will be.

Actually, the best result for the alpha wave was achieved by bior 6.8 while the worst were recorded by the haar mother wavelet. For the gamma wave, the haar mother wavelet produced the best result, while rbio 6.8 extractions were last compared to the other wavelet extractions. Then, regarding the combined best localized components of alpha and gamma, the sym 20 and coif 5 shared the first place in extracting the most accurate first five components, with the rbio 6.8 mother wavelet in the last place.

TABLE II. THE RV AVERAGE OF 5 FIRST COMPONENTS FOR THE NOETWORTHY MOTHERWAVLETS.

As an overall perception of the source localization results in evaluating our mother wavelets, we can classify the sym 20 mother wavelet as the best mother wavelet extraction over all, while the haar occupies the second place with questionable results due to our previous readings of the GOF, PSD and scalp topographies that proved the interference of frequency overlapping and noise artifacts in the sincerity of the localized components. If we eliminate the haar mother wavelet, we must crown the bior6.8 mother wavelet the second place considering the number of localized components and the best results achieved in the accuracy average of the first components in the alpha wave followed by the coif5 and sym9 mother wavelets. While On the other hand, the dmey produced a somehow moderate result in the light of the promising potential in the earlier evaluations of GOF, PSD and scalp topographies. The least favorite mother wavelet in source localization was the rbio6.8 with the worst accuracy results were recorded by all the noteworthy mother wavelets.

## IV. DISCUSSION

In this study, we have compared 51 different mother wavelets from at least 7 different families including Haar, Symlets, Daubechies, Coiflets, Discrete Meyer, Biorthogonal and reverse Biorthogonal in source localization and extraction of EEG signal. For the source localization performance comparison, the 10 mother wavelets separated from the 51 mother wavelets produced an adequate result in general. However, the sym20 outshined all the other wavelets and took the lead almost in every evaluation followed by a notable performance from bior 6.8, coif5 and sym 9, respectively. Then, the least rated results were produced by the haar and rbio6.8 mother wavelets. As a conclusion, we can say that the Symlet family generates the top results for EEG signal, as demonstrated by earlier studies with the revelation that the increase of the mother wavelets order improves the results. Then, bior6.8 and coif5 are also important mother wavelets for source localization.

Moreover, the evaluation methods included the Goodness Of Fit (GOF), the Power Spectral Density and scalp topographies in the extraction of EEG frequency sub-bands and finally source localization with the number of located sources and accuracy of localization as evaluation benchmarks. The source localization is produced via Stationary Wavelet Transform (SWT) and an Independent Component Analysis (ICA) feature extraction followed by Boundary Element Model (BEM) and Equivalent Current Dipole (ECD) solutions for the forward and inverse problem. Future studies and advancements could explore the improvement of the source localization feature extraction or forward and inverse problem solutions.

### AVAILABILITY OF DATA AND MATERIALS

The used data set of this study is available from the corresponding author upon request.

### COMPETING INTERESTS

The authors declare that there is no conflict of interest regarding the publication of this paper.

### ACKNOWLEDGMENT

| wavelet | Alpha RV average of 5 first components | Gamma RV average of 5 first components | Combined RV average of the best 5 first components |
|---|---|---|---|
| original dataset | 10.8 | | |
| sym4 | 8.9 | 9.5 | 8.6 |
| sym6 | 7.8 | 9.6 | 7.7 |
| sym9 | 6.8 | 10.2 | 6.8 |
| sym20 | 9.4 | 9.6 | 6.2 |
| haar | 10.9 | 8.3 | 6.7 |
| db5 | 7.8 | 9.7 | 7.8 |
| coif5 | 8.8 | 10.2 | 6.2 |
| bior6.8 | 6.7 | 9.6 | 6.6 |
| rbio6.8 | 9.3 | 10.4 | 9.3 |
| demy | 9.4 | 9.6 | 6.8 |

We would like to thank Dr. Nawel Jmail and Dr. Abir Hadriche for their guidance and support during the development of this project.